\documentclass[%
 reprint,
floatfix,
superscriptaddress,
showpacs,preprintnumbers,
 amsmath,amssymb,
 aps,
pra,
]{revtex4-2}
\DeclareUnicodeCharacter{2212}{-}
\usepackage{xcolor}
\usepackage{times}
\usepackage{amssymb}
\usepackage{dsfont}
\usepackage{graphicx}
\usepackage{dcolumn}
\usepackage{bm}
\usepackage{epstopdf}
\usepackage{lineno}
\usepackage[colorlinks]{hyperref}
\hypersetup{colorlinks = true,
linkcolor = blue,
filecolor = blue,
urlcolor = blue,
citecolor = blue}

\begin{document}

\title{Emergent superconductivity upon disordering a topological insulator}
\author{Carlos Eduardo S. P. Corsino}
\email{carlos.corsino@dipc.org}
\affiliation{Instituto de F\'{i}sica, Universidade Federal de Goi\'{a}s, 74.690-900, Goi\^{a}nia-GO, Brazil}
\affiliation{Donostia International Physics Center (DIPC), 20018 Donostia-San Sebastian, Spain}

\author{Hermann Freire}
\email{hermann\_freire@ufg.br}
\affiliation{Instituto de F\'{i}sica, Universidade Federal de Goi\'{a}s, 74.690-900, Goi\^{a}nia-GO, Brazil}

\author{Anurag Banerjee}
\email{anurag.banerjee@ipht.fr}
\affiliation{Universit\'{e} Paris-Saclay, Institut de Physique Th\'eorique, CEA, CNRS, F-91191 Gif-sur-Yvette, France}

\begin{abstract}
We study the emergence of superconductivity in a quantum spin Hall insulator and identify a disorder-driven enhancement of pairing arising from quantum geometry. Using sign-problem-free quantum Monte Carlo simulations of the attractive Bernevig-Hughes-Zhang (BHZ) Hubbard model, we obtain a quantum phase transition as a function of interaction strength for different impurity densities. In the clean limit, the system develops bulk superconductivity for Hubbard interaction $\vert U \vert$ above a finite critical strength. Interestingly, strong impurities significantly reduce such $\vert U \vert$ required for the onset of superconductivity. Our calculations indicate that Cooper pairing first nucleates in subgap ring states surrounding the impurities and then evolves into a globally coherent superconducting phase. Our results demonstrate that impurity-generated bound states can promote superconductivity in systems with strong quantum geometry. This mechanism is expected to be relevant in nearly flat-band systems like moir\'e materials where quantum geometry plays a dominant role.
\end{abstract}
\maketitle
\section{Introduction}
Quantum materials manifest macroscopic quantum features driven by electronic correlations~\cite{QM_Physics,Andrei2021,RMP_Cuprates_Room} or the quantum geometry~\cite{Torma_rev22,TBG_DsPRL,TopoDsTBG,SuperfluidTBG} of the underlying electronic states. While the former leads to a plethora of broken symmetry states such as superconductivity~\cite{Bednorz1986,Cao2018,UTe2_JP,Nickel2021}, charge density waves~\cite{Cava0,Park2026,Abbamonte} (CDW), among others~\cite{Kong2025,PDWNbSe2,XieLoop}, the latter can lead to topological band structures protected by symmetries~\cite{KaneMele,BernevigQSH}. The quantum metric and the Berry curvature together constitute the quantum geometric tensor, with the metric corresponding to its symmetric component and the Berry curvature to its antisymmetric component~\cite{BernRevQG,VermaRev}. The Berry curvature determines the topological character of the occupied bands, and the quantum metric quantifies the degree of orbital mixing within the Bloch wavefunctions. Unlike the Berry curvature, the quantum metric can remain finite even in topologically trivial systems~\cite{BernRevQG,VermaRev}. Consequently, quantum geometry can influence a wider range of phenomena in correlated materials, including superconductivity in twisted bilayer graphene~\cite{WhereQG,TBG_DsPRL,TopoDsTBG,SuperfluidTBG,Andrei2021,Cao2018} and other moir\'e systems, where narrow electronic bands suppress kinetic energy and enhance interaction effects.

In nearly flat-band systems, quantum geometry provides a route to superconductivity independent of the electronic dispersion~\cite{Torma_rev22,WhereQG,Peotta2015,GeometricLieb_PRL,Torma_PRB,GL_PRL_flat}. Recent studies have shown that the superfluid stiffness of isolated flat bands is fundamentally constrained by the quantum geometry of the underlying Bloch states~\cite{NishchhalPNAS,GeometricLieb_PRL,TBG_DsPRL,TopoDsTBG,SuperfluidTBG}. However, in dispersive systems, geometric contributions are often too small to be detected experimentally~\cite{Tian2023, WhereQG,Torma_PRB}. Moreover, flat bands generally support competing ordered phases~\cite{SU2AHM_flat,DebanjanPRB}, complicating the identification of purely geometric effects on superconductivity~\cite{PDW_PRL,SC_QMC}.

A promising approach for probing quantum geometry is through the response to impurities~\cite{Park2026,ABS_GraPRL,KMImpPRB}. Recent work has shown that strong impurities in systems with large quantum geometry can generate localized subgap states with ring-like spatial profiles around the impurity~\cite{wang2012impurity,sau2013bound,shan2011vacancy,slager2015impurity,PotthoffPRB,diop2020impurity,mashkoori17impurity,TMBI}, even in topologically trivial bands~\cite{queiroz2024ring,PangburnRing}. While disorder is generally expected to suppress superconductivity~\cite{RTS,GTR, DisorderDs,SambandamurthyPRL} and eventually drive a superconductor-insulator transition~\cite{GTR,Ovadia2013}, these impurity-induced states can locally enhance the low-energy density of states and provide favorable locations for Cooper pairing~\cite{KMHubbard_BdG,EdgePRL,EdgeAssaad}.

Motivated by these developments, we investigate superconductivity in a quantum spin Hall insulator~\cite{KaneMele,BernevigQSH,KM_HubbardQMC} with strong impurities. We show that impurity-induced ring states promote local pairing and enhance superconductivity by reducing the modulus of the attractive critical interaction strength~\cite{Torma_PRB} for ordering. Since ring states require strong quantum geometry irrespective of their topological character, it provides a potential route for detecting the influence of quantum geometry~\cite{xie2026,yan2026many} in superconducting materials.

\section{Model}
We consider the Bernevig-Hughes-Zhang (BHZ) model~\cite{bernevig2006quantum} on a square lattice of linear dimension $L$, which is similar to the Kane-Mele model~\cite{KaneMele,KMHubbard_BdG}: 
\begin{align}
\mathcal{H}_{\rm BHZ}&= -t\sum_{i} \left(\mathbf{\Psi}^\dagger_i  (\hat{\tau}_z  - i \hat{\tau}_x) \otimes \hat{\sigma}_z \mathbf{\Psi}_{i+\hat{x}}  \right. \nonumber \\
 &\left. + \mathbf{\Psi}^\dagger_i  (\hat{\tau}_z  + i \hat{\tau}_y )\otimes \hat{\sigma}_0 \mathbf{\Psi}_{i+\hat{y}} + \mathrm{h.c.}\right) \\
 &+ M \sum_i \mathbf{\Psi}^\dagger_i (\hat{\tau}_z \otimes \hat{\sigma}_0 ) \mathbf{\Psi}_i-\mu\sum_i \mathbf{\Psi}^\dagger_i (\hat{\tau}_0 \otimes \hat{\sigma}_0 ) \mathbf{\Psi}_i .\nonumber 
\label{eq:BHZ_model}
\end{align}
Here, $\mathbf{\Psi}_i=\left(c_{it\uparrow},c_{ib\uparrow}, c_{it\downarrow},c_{ib\downarrow}\right)^T$, where $c^{\dagger}_{i\alpha\sigma}$ ($c_{i\alpha\sigma}$) creates (annihilates) an electron at site $i$, orbital (or layer) index $\alpha=t,b$, and spin $\sigma=\uparrow,\downarrow$. The matrices $\hat{\tau}_i$ and $\hat{\sigma}_i$ denote the Pauli matrices in the orbital and spin subspaces, respectively. The chemical potential $\mu$ is chosen such that the system remains at half filling. Here, the hopping parameter $t$ determines the overall energy scale, and we set $t=1$. The parameter $M$ represents the orbital polarization and controls the topological character of the insulating state.

The BHZ model ensures that the Hamiltonian preserves time-reversal symmetry and $U(1)$ spin symmetry. The topology in this case is described by a $\mathds{Z}_2$ invariant~\cite{bernevig2006quantum,KaneMele}. If $-4t<M<4t$, the occupied bands have finite winding and lead to a quantum spin Hall insulator with gap closing at $M=0$. We focus on the parameter regime in which the noninteracting system is topologically nontrivial. To investigate the emergence of superconductivity, we supplement the noninteracting BHZ Hamiltonian with an on-site attractive Hubbard interaction~\cite{PhaseAHM,Moreo,RTS}
\begin{align}
\mathcal{H}_{U}=\sum\limits_{i,\alpha} U_\alpha (\hat{n}_{i\alpha\uparrow}-1/2)(\hat{n}_{i\alpha\downarrow}-1/2),
\end{align}
where $\hat{n}_{i\alpha\sigma} = c^{\dagger}_{i\alpha\sigma} c_{i\alpha\sigma}$. Throughout this work, we consider the symmetric attractive interactions $U_t=U_b\equiv U<0$.

In the attractive Hubbard model on a single-band square lattice at half filling, the $s$-wave superconductivity (SC) is exactly degenerate with a checkerboard charge density wave (CDW) due to a pseudospin SU(2) symmetry~\cite{Moreo,AnuragPRB}. Away from half filling, this degeneracy is lifted, and SC becomes the dominant instability, whereas the CDW is suppressed~\cite{Moreo,DosSantos,RTS,Fontenele_2022,Fontenele_2026}.

In the BHZ attractive Hubbard model, the orbital polarization parameter $M$ generates an imbalance between the orbital occupations. Consequently, the perfect nesting of the Fermi surface at half-filling is absent for $M\neq0$, suppressing the CDW order, while SC remains strong. We focus primarily on the superconducting correlation channel and analyze the CDW fluctuations and their interplay with superconductivity in Appendix~\ref{App:CDW}.

To investigate the effects of inhomogeneity, we introduce an orbital-dependent impurity potential
    ${\mathcal{H}_V=\sum_{i\alpha\sigma} V_{i\alpha} c^{\dagger}_{i\alpha\sigma} c_{i\alpha \sigma}}$.
The form of $V_{i\alpha}$ denotes the local impurity potential acting on layer (orbital) $\alpha$ at lattice site $i$. The spatial profile of $V_{i\alpha}$ determines the spatial extent of the impurity states. Here, we focus on strong impurities that have equal magnitude but opposite signs on the two orbitals. The impurity sites are periodically arranged to form a square superlattice with spacing $l$.
The resulting impurity lattice introduces an impurity concentration of $f=1/l^2$. The qualitative features are not sensitive to the specific impurity arrangement and persist for randomly distributed impurities with a similar average concentration.

For comparison, we also consider a trivial two-orbital model obtained by replacing the interorbital hopping of the BHZ Hamiltonian with a conventional interorbital hybridization. The corresponding noninteracting Hamiltonian is given by
\begin{align}
\mathcal{H}_{\rm t}&= -t \sum_{\langle i,j\rangle,\alpha\sigma} c^\dagger_{i\alpha\sigma} c_{j\alpha\sigma} -t_\perp \sum_{i\sigma} c^{\dagger}_{it\sigma} c_{ib\sigma} + \mathrm{h.c.} \nonumber \\ &+ M\sum_{i\sigma} (c^{\dagger}_{it\sigma} c_{ib\sigma} - c^{\dagger}_{ib\sigma} c_{it\sigma} ),
\label{Eq:Trivial}
\end{align}
where $t_{\perp}$ denotes the interorbital (interlayer) hybridization, and $M$ is the orbital polarization parameter controlling the electron density on the two orbitals. For a finite $M$, we avoid perfect Fermi surface nesting and thus the interplay with CDW ordering for attractive interactions.  
\section{Method}
To solve the BHZ-Hubbard Hamiltonian, we utilize the numerically exact method of auxiliary-field quantum Monte Carlo (AFQMC) simulations~\cite{Sugar}, as implemented in the ALF (Algorithms for Lattice Fermions) package~\cite{ALFcode}. Here, the interaction term is decoupled through a Hubbard-Stratonovich transformation, converting the interacting fermion problem into an ensemble of noninteracting fermions coupled to fluctuating auxiliary fields~\cite{Sugar}. Physical observables are then obtained by stochastically sampling these auxiliary-field configurations. The attractive BHZ-Hubbard interaction remains free of the exponential fermionic sign problem at low temperatures~\cite{Moreo,SC_QMC,RTS,DosSantos}.

We implemented the BHZ-Hubbard Hamiltonian within the ALF Hubbard class~\cite{ALFcode} and carried out both finite-temperature (see Appendix~\ref{App:FiniteT}) and ground-state simulations. The results presented in the main text are obtained using the projective AFQMC algorithm, which targets ground-state properties. The central idea is to isolate the many-body ground state through imaginary-time evolution. Starting from a suitably chosen trial wave function that possesses a finite overlap with the true ground state, repeated propagation in imaginary time exponentially suppresses contributions from excited states. As the projection time is increased, the wave function evolves toward the exact ground state of the interacting Hamiltonian, allowing expectation values of physical observables to be evaluated. We measure observables only after sufficiently long projection times.

\section{Results}
To characterize the emergence of superconductivity, we compute the equal-time $s$-wave pairing correlations using AFQMC simulations. The orbital resolved pair-pair correlation function~\cite{Moreo,PhaseAHM,DosSantos} is defined as
\begin{align}
    P_{\alpha\beta}(i,j) = \langle c^\dagger_{i\alpha\uparrow} c^\dagger_{i\alpha\downarrow} c_{j\beta\downarrow} c_{j\beta\uparrow} + \mathrm{h.c.} \rangle.
\end{align}
We extract $\tilde{P}_{\alpha\beta}(\mathbf{q})$ by the standard Fourier transform.
We define
$P_{\alpha\beta} \equiv \tilde{P}_{\alpha\beta}(\mathbf{q}=0)$
as the measure of long-range superconducting order. For the symmetric interaction considered in this work, $U_t=U_b\equiv U$, yielding $P_{tt}=P_{bb}=P$. 

We extract the SC correlation length from the momentum dependence of the pairing structure factor, following the standard second-moment estimator~\cite{SandvikLec,SC_QMC}
\begin{align}
    \xi_a=\frac{L}{2 \pi} \sqrt{\frac{\widetilde{P}(0)}{\widetilde{P}(\mathbf{q}_1)}-1}\hspace{0.05cm},
\end{align}
where $\mathbf{q}_1=(2\pi/L,0)$ is the smallest nonzero momentum allowed by the finite lattice. Near a continuous SC transition, $\xi_a$ enables an accurate determination of the critical interaction strength. In addition to the superconducting correlations and correlation length, we monitor the average electron density in each orbital. 

The AFQMC simulations are performed on square lattices with linear dimensions ranging from $L=8$ to $L=20$ with periodic boundary conditions. For the BHZ-Hubbard model, we set the value of $\mu=0$ such that the system is half-filled. Later, in the main text, we present the results for $M=0.1$, while the results for $M=0.3$ are presented in Appendix~\ref{App:M0_3}.
  
 \begin{figure}[t]
    \centering
    \includegraphics[width=1.0\linewidth]{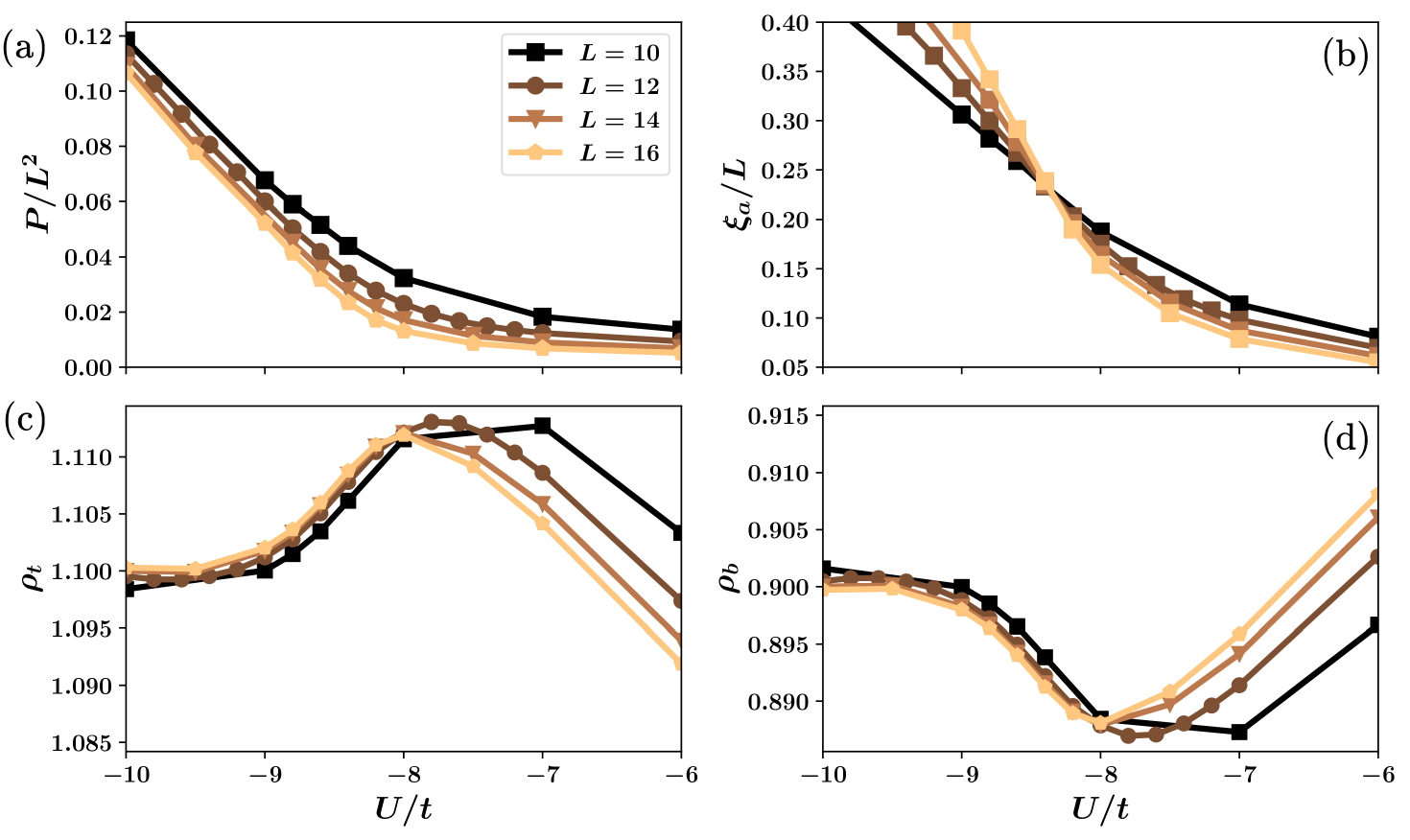}
    \caption{Results from AFQMC simulations for the BHZ-Hubbard model without impurity. (a) Pairing correlations as a function of $U$ for different system sizes. Beyond a critical value of $\vert U \vert$, $s$-wave SC emerges in the bulk of the topological insulator. (b) Pairing correlation length as a function of interaction strength for different system sizes. The crossing point identifies the critical $U_c$. (c) and (d) Average electron density for the top and bottom layers, respectively. Near the critical $U$, the electron density reaches a maximum in the top layer and a minimum in the bottom layer.
}
    \label{fig:fig1}
\end{figure}
\subsection{Clean system}

Since the system is a gapped topological insulator, superconductivity does not develop at weak attractive interactions. Similar behavior has been reported for  honeycomb lattice and related topological band insulators~\cite{UchoaPRL,DQMC_AHM_Honey,BdG_PRB_graphene_disorder,KM_HubbardQMC,EdgePRL,KMHubbard_BdG}. As shown in Fig.~\ref{fig:fig1}(a), the pairing correlations remain small at weak coupling $\vert U \vert$ and increase rapidly beyond a critical interaction strength $\vert U_c \vert$. The weak size dependence of the pairing structure factor for $L=10$ to $L=16$ indicates the onset of long-range superconducting order.

A more accurate estimate of the transition is obtained from the superconducting correlation length shown in Fig.~\ref{fig:fig1}(b). In the superconducting phase, the correlation length grows with the system size, whereas it remains finite in the insulating phase. The crossing of $\xi/L$ for different system sizes identifies the quantum critical point, yielding a critical interaction strength of $U_c \approx -8.5t$.

The transition is also reflected in the orbital-resolved electron densities. As shown in Figs.~\ref{fig:fig1}(c) and \ref{fig:fig1}(d), the electron density in the two orbitals exhibits an extremum near $U_c$, with electrons transferring from one orbital to the other as the modulus of the interaction strength increases. The location of this extremum closely tracks the superconducting transition and provides an additional signature of the onset of bulk superconductivity.

\begin{figure}[t]
    \centering
    \includegraphics[width=1.0\linewidth]{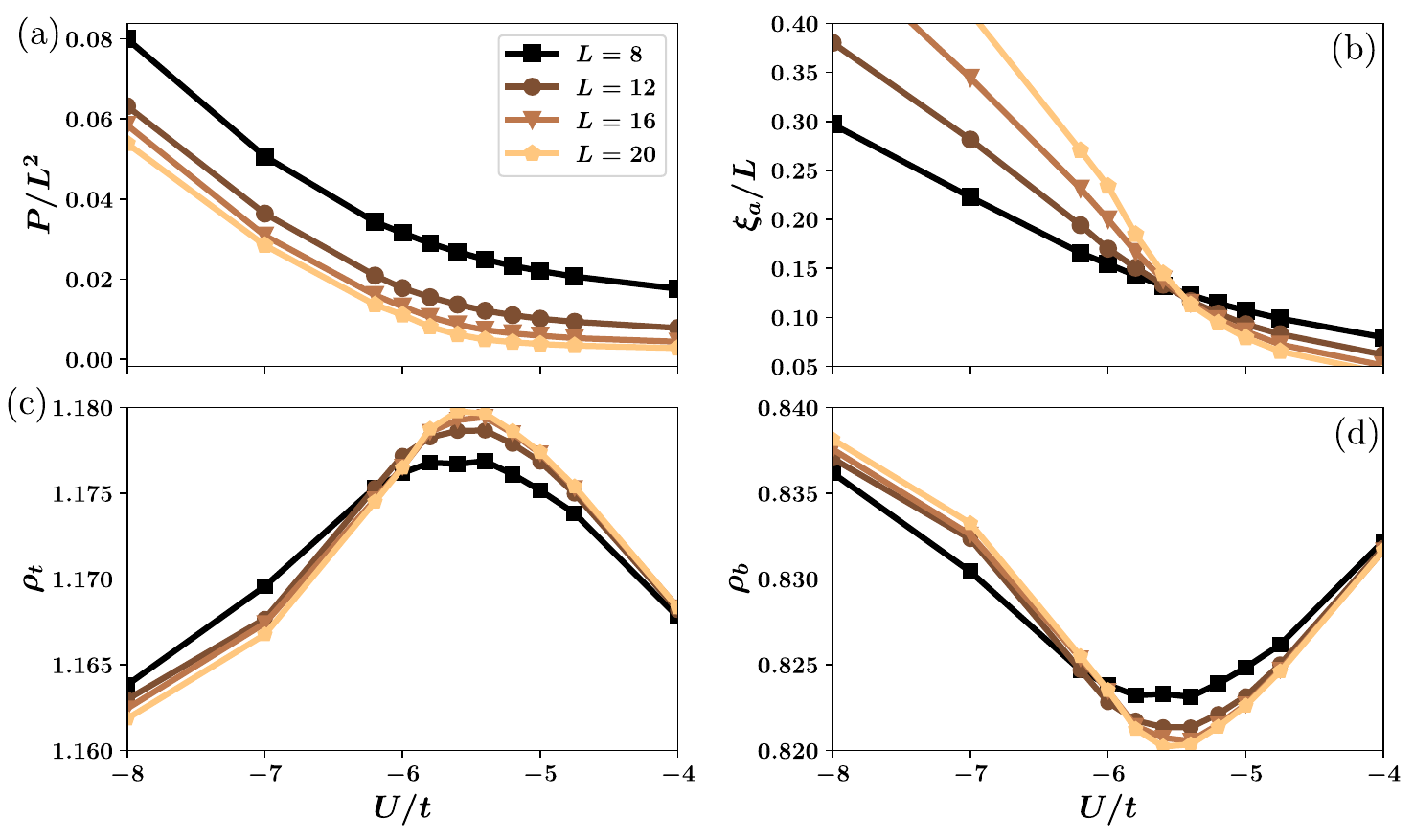}
    \caption{Results from AFQMC simulations for the BHZ-Hubbard model with impurity density $f=1/16$ (a) Pairing correlations as a function of $U$ for different system sizes. (b) Pairing correlation length as a function of interaction strength for different system sizes, where the crossing point is $U_c$. The modulus of the critical $\vert U_c \vert$ for the disordered system is significantly lower than that of the clean system.  (c) and (d) Average electron density for the top and bottom layers, respectively. Near the critical $U$, the electron density reaches a maximum in the top layer and a minimum in the bottom layer.}
    \label{fig:fig2}
\end{figure}
\subsection{Inhomogeneous system}
Next, we study a system with a spatially periodic impurities of strength $V_0 = -40t$ in the top layer and $V_0 =40t$ in the bottom layer. Impurities of equal magnitude but opposite sign are chosen to preserve the overall particle-hole symmetry of the system; however, our conclusions remain valid for other choices of impurity polarity on the two orbitals. In Fig.~\ref{fig:fig2}(a), we present results for an impurity density $f = 1/16$, using system sizes commensurate with the impurity periodicity.

The pairing correlations grow at a value of $\vert U \vert$ that is substantially lower than in the clean case. Analyzing the crossing of the correlation length curves in Fig.~\ref{fig:fig2}(b) yields $U_c \approx -5.5 t$, whose absolute value is significantly smaller than the critical coupling of the clean system.

We attribute this reduction to the formation of sub-gap ring states in the bulk in the vicinity of each impurity site, which act as nucleation centers for Cooper pairing. Once the Cooper pairs localized around neighboring impurities acquire long-range phase coherence, the entire bulk of the system becomes superconducting.

Figures~\ref{fig:fig2}(c) and \ref{fig:fig2}(d) display the average electron density in the top and bottom layers, respectively. Consistent with the behavior observed in the clean system, the layer density exhibits a maximum in the top layer and a minimum in the bottom layer near $\vert U_c \vert$, thereby providing a strong experimental signature of the impurity-induced bulk superconducting transition.

\begin{figure}[h!]
    \centering
    \includegraphics[width=1.0\linewidth]{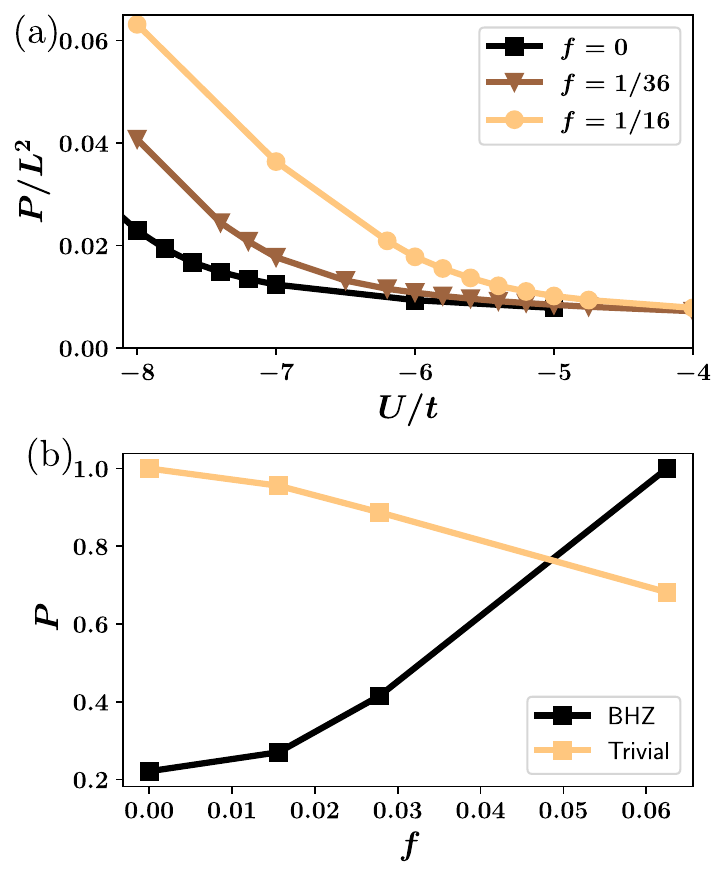}
    \caption{(a) Dependence of the superconducting correlations on the interaction strength $U$ for different impurity densities $f$ (where $f=0$ is the impurity free system). (b) Superconducting correlations as a function of impurity density $f$ for systems with topological bands at $U=-6t$ and trivial bands $U=-4t$. In the topological phase, the superconducting correlations are enhanced with increasing impurity density, whereas in the trivial bands they are reduced with increasing impurity density.}
    \label{fig:fig3}
\end{figure}

The emergence of superconductivity induced by disorder in the topological phase is further illustrated in Fig.~\ref{fig:fig3}. As shown in Fig.~\ref{fig:fig3}(a), the pairing correlations begin to increase for $\vert U \vert$ beyond a critical interaction strength, whose value depends strongly on the impurity concentration. Remarkably, $\vert U_c \vert$ decreases systematically with increasing impurity density, indicating that disorder promotes the onset of superconductivity in the topological insulator. 

In Fig.~\ref{fig:fig3}(b), we compare the normalized superconducting correlations at $U=-6t$ for the BHZ-Hubbard model with those of the $U=-4t$ trivial two-band model introduced in Eq.~(\ref{Eq:Trivial}). The two systems exhibit qualitatively distinct responses to increasing impurity density. For the trivial model, the superconducting correlations decrease monotonically with disorder. For trivial insulators, strong impurity scattering locally suppresses the pairing amplitude and disrupts phase coherence.

In contrast, the topological system exhibits an enhancement of superconducting correlations with increasing impurity density. The strong orbital mixing generates impurity-induced subgap states that substantially increase the low-energy spectral weight available for pairing. Rather than acting solely as pair-breaking centers, the impurities create electronic states that promote Cooper-pair formation. As a result, disorder enhances superconductivity in the topological band structures, in stark contrast to its role in trivial bandstructures.

\subsection{Phase diagram}
\begin{figure}[t]
    \centering
    \includegraphics[width=1.0\linewidth]{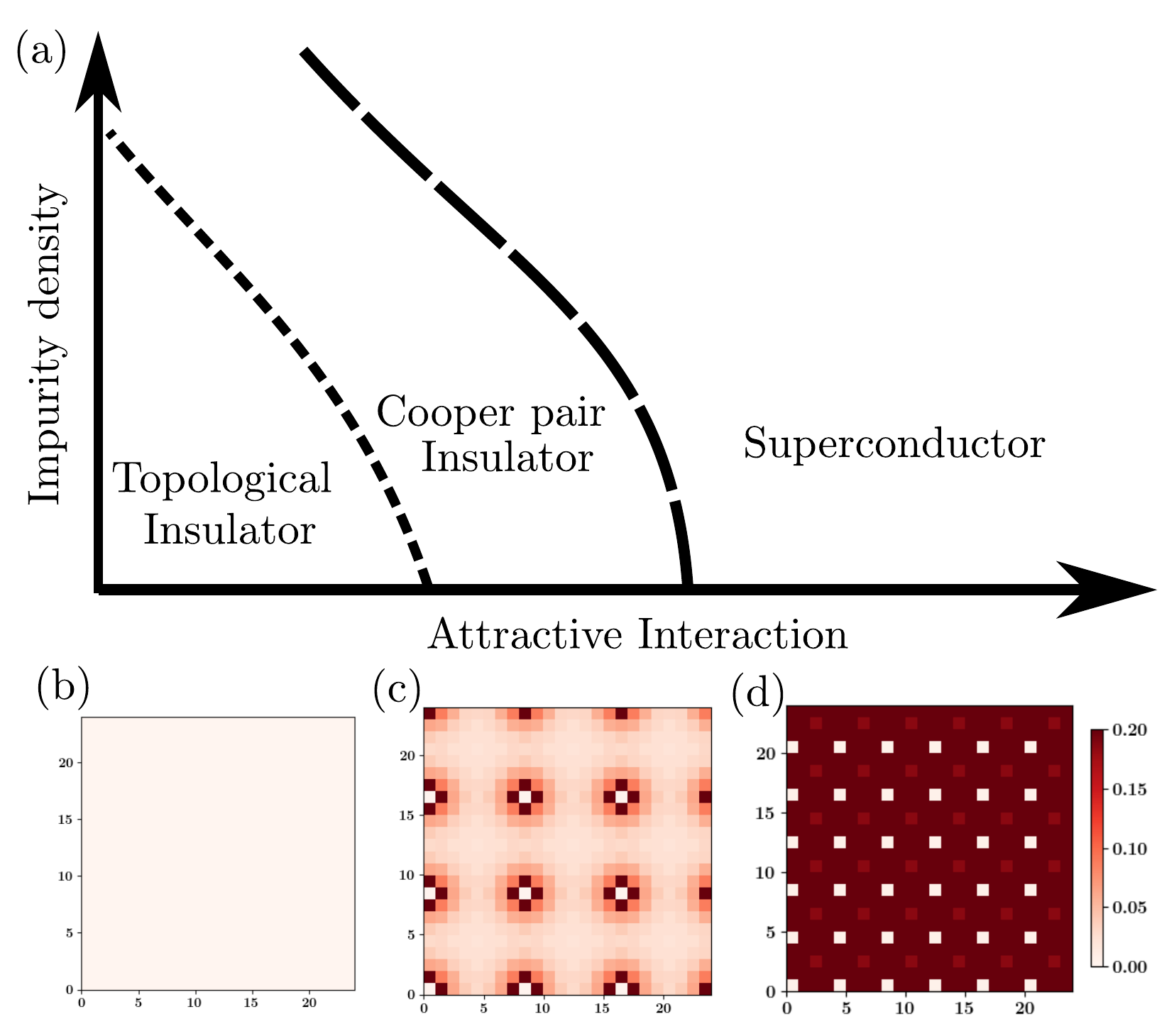}
    \caption{(a) Schematic phase diagram of the attractive BHZ-Hubbard model in the interaction strength and impurity density plane. In the clean limit, the system undergoes a transition from a quantum spin Hall insulator to a superconducting phase at a critical interaction strength $\vert U_c \vert$. Increasing the impurity density systematically lowers $\vert U_c \vert$. Self-consistent BdG calculations further suggest the emergence of a Cooper-pair insulating regime, in which local pairing develops around impurity-induced subgap states before the establishment of global phase coherence. (b) Spatial profile of the pairing amplitude in the topological insulator with SC pairing absent. (c) Same for Cooper-pair insulating phase, characterized by localized pairing around impurities. (d) Superconducting phase where the pairing is only suppressed on the impurities.
 }
    \label{fig:fig5}
\end{figure}

The disorder-enhanced superconductivity observed in the AFQMC simulations is also captured within the Bogoliubov-de Gennes (BdG) calculations, as shown in Appendix~\ref{App:BdG}. As illustrated schematically in Fig.~\ref{fig:fig5}(a), the clean system undergoes a transition from a topological insulating phase to a bulk superconducting phase when the absolute value of the attractive interaction exceeds a critical value $\vert U_c \vert$. The introduction of strong impurities substantially lowers the modulus of this critical interaction strength, and superconductivity can be stabilized at interaction strengths that are insufficient to induce pairing in the clean system. 

Further insight into the nature of this transition is obtained from the real-space superconducting order parameter calculated within the self-consistent BdG framework. Representative pairing profiles are shown in Figs.~\ref{fig:fig5}(b)-\ref{fig:fig5}(d). In the weak-coupling regime, $\vert U \vert<\vert U_c \vert$, the system remains in the topological insulating phase, and no superconducting pairing develops, even in the vicinity of the impurity-induced bound states [Fig.~\ref{fig:fig5}(b)].

For the modulus of interaction strengths well above $\vert U_c \vert$, the superconducting order parameter extends throughout the system, as shown in Fig.~\ref{fig:fig5}(d). In this regime, long-range coherence is established across the sample, producing a bulk superconducting state. The superconducting order is only suppressed at the impurity sites, where the strong local potential prohibits pairing.

Between these two limits lies an intermediate state illustrated in Fig.~\ref{fig:fig5}(c) for $U=-3.8t$ and $f=1/64$. Here, substantial pairing develops around the impurity-induced ring-like bound states, indicating the formation of local Cooper pairs. However, the pairing amplitude decays rapidly away from the impurities and remains negligible in the regions separating neighboring impurity sites. The locally paired regions remain disconnected and fail to establish global coherence. 

The existence of such an intermediate regime highlights the importance of impurity-induced subgap states in the pairing mechanism. Superconductivity first nucleates in regions where the low-energy spectral weight is enhanced by the impurities and only evolves into a coherent superconducting phase once the corresponding paired regions begin to overlap. The resulting local pairing can emerge for absolute values of the interaction strengths significantly below those required for bulk superconductivity and should gap out the edge modes before global superconducting coherence is established.

\section{Discussion}
In this work, we studied the emergence of superconductivity in a quantum spin Hall insulator described by the attractive BHZ-Hubbard model. Using numerically exact AFQMC simulations, we established that strong impurities enhance superconductivity by reducing the absolute value of the critical interaction required for the onset of a long-range order. This behavior contrasts with the conventional expectation that disorder suppresses superconductivity~\cite{Moreo,GTR}. Therefore, the identification of an enhancement of superconductivity with disorder provides strong evidence of superconducting systems with substantial quantum geometric effects.

This enhancement originates from impurity-induced subgap states, where strong impurities create bound states within the insulating gap~\cite{Yazdani_STM_ring,Xu3DRing,STM_ring}. These states provide favorable locations for Cooper pairing. As the modulus of the interaction strength increases, the pairing first develops in these impurity-induced ring states and subsequently overlaps with each other to spread throughout the system. This picture is supported by the real-space BdG solutions, which reveal an intermediate regime characterized by strong local pairing around impurities but weak pairing in the regions separating them. Similar disorder-induced enhancement of SC correlations is also observed in gapless Dirac fermions~\cite{BdG_PRB_graphene_disorder}.

Recent studies have shown that impurity-induced ring states can arise not only in topological insulators but also in systems where quantum geometry plays a dominant role~\cite{queiroz2024ring,PangburnRing}.  Therefore, such a mechanism extends to trivial insulators with strong quantum geometry. In particular, nearly flat-band systems, as expected for moiré materials, can support such localized subgap states even in the absence of a topological band inversion. The enhancement of superconductivity through impurity generated low-energy states can therefore extend to a broader class of quantum materials where geometric effects dominate single-particle dispersion.

Our results highlight the limitations of topological classification of superconducting phases solely within non-self-consistent mean-field Hamiltonians. The boundary modes  become unstable to ordering for lower absolute interaction strength than those required for bulk ordering. As a result, local orders can emerge near topological boundaries, while the bulk remains insulating. 

Our findings can be relevant for topological insulator-superconductor heterostructures~\cite{Bismuthene_QSH,RothHgTe,PRLQW}. We predict that superconductivity can preferentially nucleate at defects, interfaces, and impurity-induced boundary modes before extending into the bulk. Such inhomogeneous pairing should be accessible through scanning tunneling microscopy.

\section*{Acknowledgments}
The authors acknowledge F. Assaad for providing the open-source ALF package and S. Biswas for helpful discussions regarding the implementation of the AFQMC simulations. The authors thank E. Pangburn, C. P\'{e}pin and I. Froldi for helpful discussions. The numerical calculations were performed on the Kanta cluster at IPhT and at the Laborat\'{o}rio Multiusu\'{a}rio de Computa\c{c}\~{a}o de Alto Desempenho (LaMCAD) of UFG in Goi\^{a}nia-GO. H.F. acknowledges funding from the Conselho Nacional de Desenvolvimento Cient\'{i}fico e Tecnol\'{o}gico (CNPq) under Grants No. 305575/2025-2, 404274/2023-4, and 407658/2025-4. H.F. also  acknowledges the support of the INCT project Advanced Quantum Materials, involving the Brazilian agencies CNPq (Proc. 408766/2024-7), FAPESP (Proc. 2025/27091-3), and CAPES.

\appendix
\section{Bogoliubov-de Gennes (BdG) results}
\label{App:BdG}
\begin{figure}[h!]
    \centering
    \includegraphics[width=1.0\linewidth]{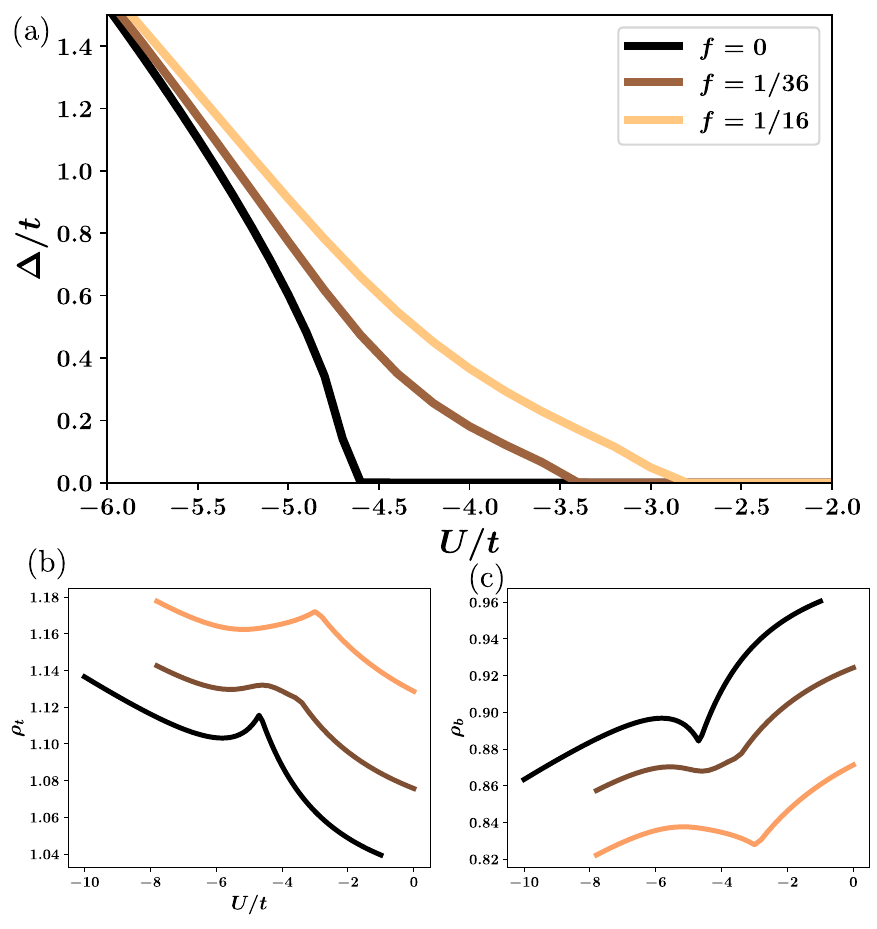}
    \caption{(a) Mean-field SC pairing amplitude as a function of  $U$ for different impurity concentrations. As the impurity concentration increases, the absolute value of the critical interaction strength $\vert U_c \vert$, required for the onset of superconductivity, is significantly reduced compared to the clean system. (b) Average electron density in the top layer as a function of $U$. (c) Average electron density in the bottom layer as a function of $U$. The minima in the electron density identify the critical interaction strength at which the system undergoes the transition into the superconducting phase.}
    \label{fig:fig4}
\end{figure}
To gain insight into the spatial structure of the superconducting state, we also perform self-consistent BdG calculations in the presence of strong impurities~\cite{PdGbook,GTR,GTR_PRB,Proximity}. We perform a mean-field decoupling of the interaction term in both pairing and Hartree channels. This introduces a local superconducting pairing $\Delta_{\alpha i} =\langle c_{i \alpha \downarrow} c_{i\alpha\uparrow}\rangle $ and a site-dependent Hartree potential from the local density ${\rho_{i \alpha \sigma}= \langle c^{\dagger}_{i \alpha \sigma} c_{i \alpha \sigma} \rangle}$ that are determined self-consistently. Here, we asssume a paramagnetic solution: $\rho_{i \alpha \uparrow} =\rho_{i \alpha \downarrow} $.

The presence of impurities breaks the translational symmetry periodically. We therefore solve the BdG equations within a single impurity supercell and employ a repeated-zone (supercell) scheme to construct the full spectrum, following the methodology of Refs.~\cite{black-induced_2013,ZhuPRB_Proximity,PRBtJ_22,Proximity_Marsiglio}. The calculations are performed on lattices with linear size up to $L=120$.

The disorder-enhanced superconductivity observed in the AFQMC simulations is also captured within the BdG calculations, as shown in Fig.~\ref{fig:fig4}(a). The absolute value of the critical interaction strength $\vert U_c \vert$, obtained from the BdG calculations, is systematically smaller than that extracted from the numerically exact AFQMC simulations. Such an underestimation of the critical coupling is a well-known limitation of mean-field approaches~\cite{UchoaPRL,DQMC_AHM_Honey}. Nevertheless, the BdG calculations capture the reduction of $\vert U_c \vert$ with increasing impurity concentration, and the characteristic behavior in the average electron densities of the top and bottom layers indicating the superconducting transition.

\section{Temperature dependence of the clean system}
\label{App:FiniteT}
\begin{figure}[t]
\centering
\includegraphics[width=\linewidth]{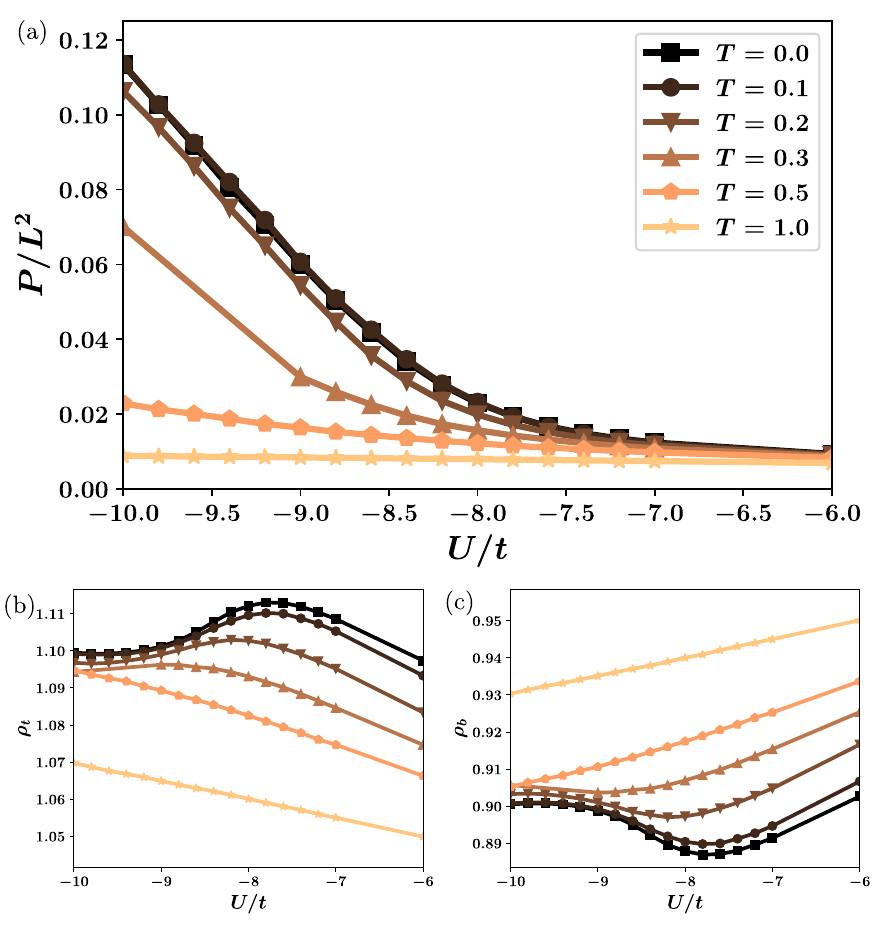}
\caption{(a) Dependence of the superconducting correlations on the interaction strength $U$ for different temperatures. (b) Average electron density in the top layer as a function of $U$ for different temperatures. (c) Average electron density in the bottom layer as a function of $U$ for different temperatures. The extrema in the electron density observed near the critical interaction strength gradually weaken with increasing temperature and disappear above $T_c$.}
\label{fig:figApp1}
\end{figure}

To characterize the finite-temperature behavior of the clean system, we study the superconducting correlations from AFQMC calculations on 2D systems of linear dimension $L=12$. As shown in Fig.~\ref{fig:figApp1}(a), increasing temperature suppresses the superconducting correlations throughout the ordered phase, reflecting the usual thermal destruction of Cooper pairs. While the magnitude of the pairing correlations decreases with temperature, the interaction strength at which superconductivity first emerges remains nearly unchanged over the temperature range considered. Above a critical temperature $T_c$, however, the superconducting correlations vanish and the system no longer develops long-range superconducting order for any value of $U$.

The temperature dependence of the orbital-resolved electron density is shown in Figs.~\ref{fig:figApp1}(b) and \ref{fig:figApp1}(c). At low temperatures, the average electron density exhibits pronounced extrema near the interaction strength associated with the onset of superconductivity. As the temperature is increased, these extrema are progressively suppressed. Upon approaching $T_c$, the extrema disappear entirely, and the density evolves into a smooth, nearly linear function of $U$. 

\begin{figure}[t]
\centering
\includegraphics[width=\columnwidth]{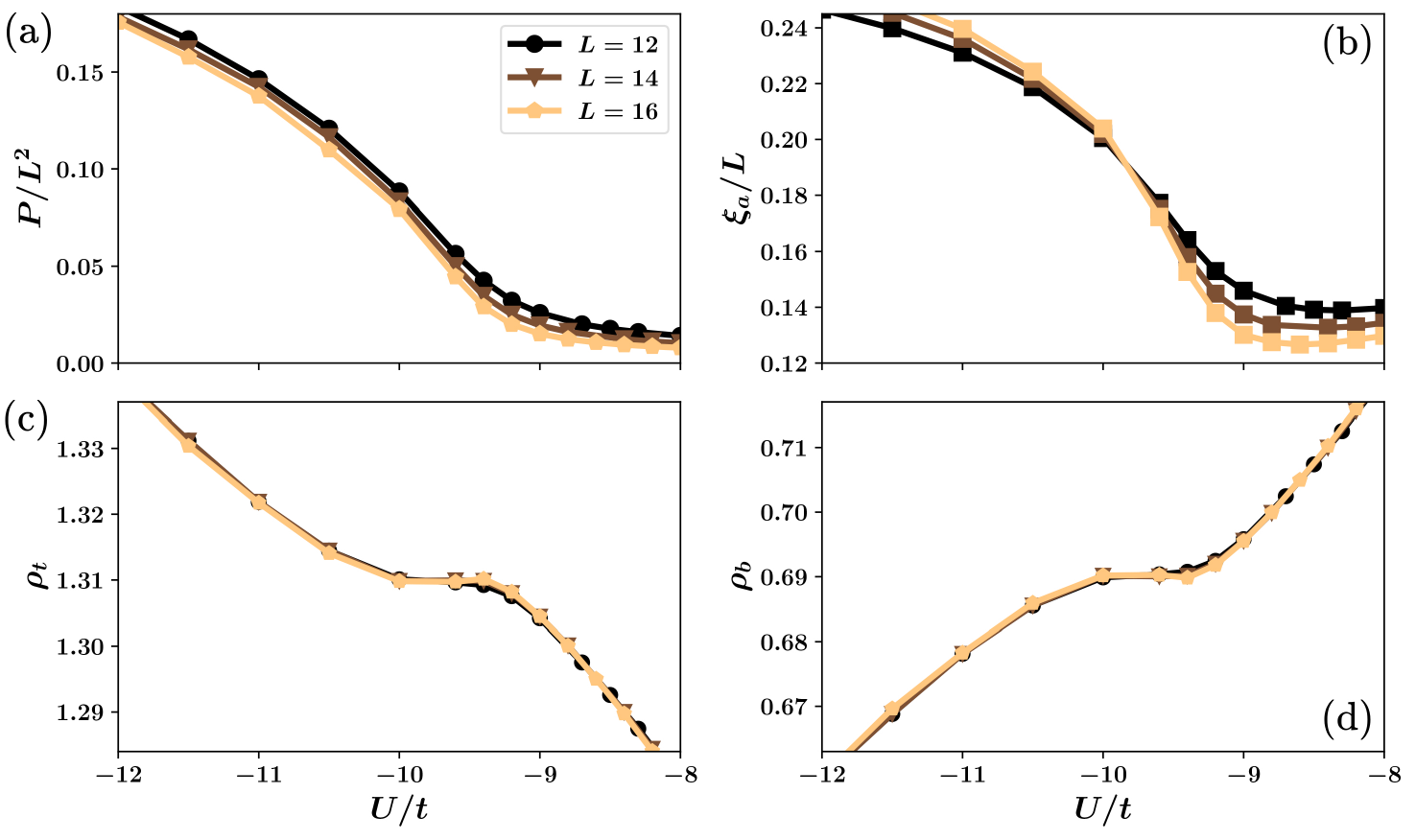}
\caption{Results from AFQMC simulations for $M=0.3$ for a clean system at zero temperature. (a) Pairing correlations as a function of $U$ for different system sizes. Beyond a critical value of $\vert U \vert$, $s$-wave SC emerges in the bulk of the topological insulator. (b) Pairing correlation length as a function of interaction strength for different system sizes. The crossing point identifies the critical $U_c \approx -9.8\,t$. (c) and (d) Average electron density for the top and bottom layers, respectively.}
\label{fig:C1}
\end{figure}

\section{Clean system for $M=0.3$}
\label{App:M0_3}
We consider the clean BHZ-Hubbard model with a larger orbital polarization, $M=0.3$, which enhances the orbital imbalance and enlarges the insulating gap, thereby suppressing the ordering tendencies. Figure~\ref{fig:C1}(a) shows the superconducting pairing correlations as a function of interaction strength. As in the $M=0.1$ case, the pairing correlations remain small at weak coupling and increase rapidly for $\vert U \vert$ beyond a critical interaction strength. However, the onset of superconductivity is shifted to larger values of $\vert U \vert$, indicating that stronger interactions are required to overcome the larger insulating gap.

A more precise estimate of the transition is obtained from the superconducting correlation length shown in Fig.~\ref{fig:C1}(b). The crossing of $\xi_a/L$ for different system sizes identifies the quantum critical point at $U_c \approx -9.8t$. The modulus of this value is significantly larger than the corresponding critical interaction for $M=0.1$, confirming that increasing the orbital polarization suppresses the superconducting instability.

The transition is also reflected in the orbital-resolved electron densities. As shown in Figs.~\ref{fig:C1}(c) and \ref{fig:C1}(d), the density of the top orbital exhibits a weak maximum near $U_c$, while the bottom orbital displays a corresponding weak minimum. Similar to the behavior observed for $M=0.1$, these extrema provide a signature of the superconducting transition.
\section{Density wave order for finite orbital polarization}
\label{App:CDW}
The CDW is characterized using equal-time density-density correlations which is defined as
\begin{align}
    S_{\alpha \beta}(i,j) = \left\langle \sum_{\sigma,\sigma^\prime} c^\dagger_{i \alpha \sigma} c_{i \alpha \sigma} c^\dagger_{j \beta \sigma^\prime} c_{j \beta \sigma^\prime}  \right\rangle.
\end{align}

\begin{figure}[b]
\centering
\includegraphics[width=\linewidth]{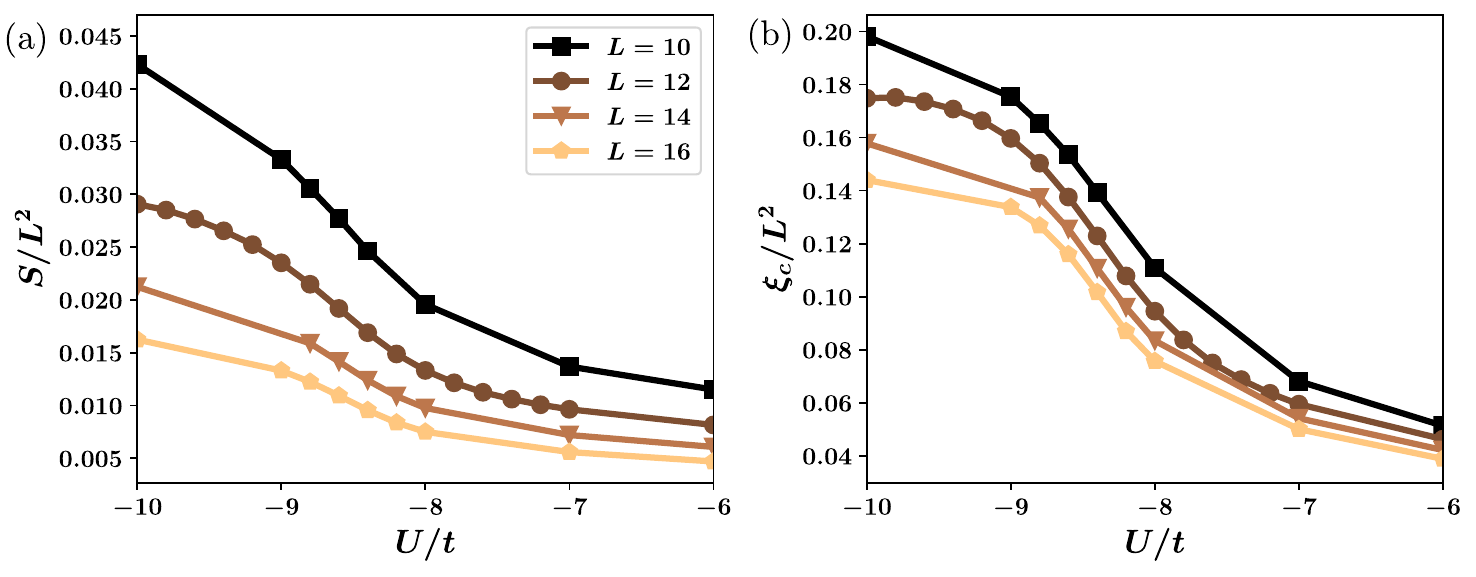}
\caption{Results from AFQMC simulations for the clean BHZ-Hubbard model with ${M}=0.1$ at zero temperature. (a) CDW structure factor at $\mathbf{Q}=(\pi,\pi)$ as a function of interaction strength $U$ for different lattice sizes. (b) Corresponding CDW correlation length as a function of $U$. The absence of finite-size scaling crossings and the lack of significant growth of the correlations provide no evidence for a CDW instability.
}
\label{fig:figApp3}
\end{figure}

We Fourier transform the real-space correlation function and obtain $\tilde{S}_{\alpha \beta}(\mathbf{q})$. The dominant CDW fluctuations occur at the checkerboard ordering wavevector $\mathbf{Q}=(\pi,\pi)$. We therefore define the CDW structure factor as
$S_{\alpha\alpha}=\tilde{S}_{\alpha\alpha}(\pi,\pi)$.
For the symmetric interaction considered here, $S_{tt}=S_{bb}=S$. 

To quantify the spatial extent of the CDW correlations, we extract the CDW correlation length using the second-moment estimator,
\begin{align}
    \xi_c=\frac{L}{2 \pi} \sqrt{\frac{\tilde{S}(\pi,\pi)}{\tilde{S}(\mathbf{q_1})}-1}\hspace{0.05cm},
\end{align}
where $\mathbf{q}_1=(\pi-2\pi/L,\pi)$ is the closest wavevector from the dominant one allowed by the finite lattice. The behavior of $S$ and $\xi_c$ as a function of system size provides a direct probe of CDW ordering tendencies in the model.

Figure~\ref{fig:figApp3}(a) shows the CDW structure factor as a function of interaction strength for several system sizes. While the CDW correlations increase for strong attractive interactions, becoming noticeable near $U\approx -8.5t$, yet the finite-size dependence remains pronounced. The growth of the structure factor does not exhibit the scaling behavior expected for the onset of long-range CDW order and thus indicate short-range charge density fluctuations.

This conclusion is further supported by the behavior of the CDW correlation length shown in Fig.~\ref{fig:figApp3}(b). Unlike the superconducting correlation length, which displays a clear enhancement near the superconducting transition, the CDW correlation length decreases with increasing system size and does not exhibit a crossing point. The absence of a scale-invariant behavior indicates that the charge correlations remain short-ranged throughout the parameter regime studied.

These results show that a finite orbital polarization suppresses the CDW instability, leaving superconductivity as the dominant ordering tendency in the BHZ attractive Hubbard model.

\bibliography{ref}
\bibliographystyle{apsrev4-2}
\end{document}